\documentclass{aa}
\usepackage{graphicx}
\def\ssim{\setbox0=\hbox{$\propto$}%
\setbox1=\hbox{$<$}\dimen0=\ht1%
\advance\dimen0by-1.2pt\,\lower.6\dimen0%
\copy0\kern-\wd0\raise.4\dimen0\copy1 \,}

\def\gsim{\setbox0=\hbox{$\propto$}%
\setbox1=\hbox{$>$}\dimen0=\ht1%
\advance\dimen0by-1.2pt\,\lower.6\dimen0%
\copy0\kern-\wd0\raise.4\dimen0\copy1\,}

\def\lambdab{\lambda\mkern-9mu\lower1.2pt\hbox{$\mathchar'26$}}%

\begin{document}
   \title{The blue to red supergiant ratio in young clusters at various metallicities}
   
   \titlerunning{The blue to red supergiant ratio}


 \author{P. Eggenberger, G. Meynet, A. Maeder}

     \institute{Geneva Observatory CH--1290 Sauverny, Switzerland\\
              email: eggenbe7@etu.unige.ch \\
              email: Georges.Meynet@obs.unige.ch\\
	      email: Andre.Maeder@obs.unige.ch     }

   \date{Received / Accepted}

\abstract{ We present new determinations of the blue to red supergiant ratio ($B/R$) in young open clusters at various 
metallicities.
For this purpose, we examine the HR diagrams of 45 clusters in the Galaxy and of 4 clusters in the Magellanic
Clouds. The identification of supergiants is based on spectroscopic measurements (with photometric counts to
check the results). The new counts confirm the increase of the $B/R$
ratio when the metallicity increases with the following normalized relation : $\frac{B/R}{(B/R)_{\odot}} \cong 0.05\cdot e^{3 \frac{Z}{Z_{\odot}}}$, where $Z_{\odot}=0.02$ and $(B/R)_{\odot}$ is the value of $B/R$ at $Z_{\odot}$ which depends on the definition of $B$ and $R$ 
and on the age interval considered (e.g. for spectroscopic counts including clusters with $\log age$ between 6.8 and
7.5, $(B/R)_{\odot}\cong3$ when $B$ includes O, B and A supergiants).
               }

   \maketitle
%

\section{Introduction}

The variation with metallicity $Z$ of the number of blue and red 
supergiants is important in relation
to the nature of the supernova progenitors in different environments (Langer 1991ab)
and its effects on the luminous star populations in galaxies
(e.g. Cervino \& Mas-Hesse \cite{ce94}; Origlia et al. \cite{or99}).
This ratio also constitutes an important and sensitive test for stellar evolution models, because it is very sensitive to mass
loss, convection and mixing processes (Langer \& Maeder \cite{lm95}). 
Stellar evolution models can 
usually, by adjustement of parameters such as the rate of mass loss by stellar winds, reproduce the observed ratio for a given metallicity, but are not able to reproduce its variation with the
metallicity. Thus, the problem of the blue to red supergiant ratio ($B/R$ ratio)
remains one of the most severe problems in stellar evolution.

Since the last studies of the $B/R$ ratio in the Galaxy and in the Magellanic Clouds 
more than fifteen years ago (Meylan \& Maeder \cite{mm82}; Humphreys \& McElroy \cite{hm84}), 
many new spectroscopic and photometric measurements have been performed.
Our aim here is to reexamine this question
accounting for these recent observational improvements. 
We chose to use stellar clusters instead of field stars for obvious reasons: stars are at the same distance, with the same age and
above all have the same chemical composition. Moreover, the knowledge of the age of the clusters enables us
to estimate the initial masses of the supergiants used to derive the $B/R$ ratio.

Thus, the present work is in the continuity of the work of
Meylan \& Maeder (\cite{mm82}). However, the present study presents two significant improvements.
First, the identification of 
supergiants is based on spectroscopic measurements instead of on photometric colours. 
Spectroscopic measurements allow a more accurate differentiation between supergiants and main sequence stars (see however
the discussion in Sect. 4).
Second, the number of stellar clusters in our galactic sample is
increased by a factor of four.

In Sect. 2, we review the various studies about the $B/R$ ratio. Section 3 presents the new counts.
Our results are discussed in Sect. 4, and conclusions are given in Sect. 5.

\section{Previous studies}

Walker (\cite{wa64}) observed that the $B/R$ ratio in M33 decreases
when the distance to the centre of the galaxy increases.
Van den Bergh (1968) explained this variation by
a radial metallicity gradient in the disk of M33, the metallicity being higher in the inner parts. 
Humphreys \& Sandage (1980) 
reexamined the $B/R$ ratio in M33 and confirmed its decrease when the galactocentric distance increases.
Freedman (1985) contested the results of Humphreys \& Sandage (\cite{hs80}) concerning the
variation of the $B/R$ ratio in M33 and concluded that this was just an effect of incompleteness.
However, Ivanov (1998) concluded that the diminution was
real and was in good agreement with the radial metallicity gradient observed in M33 (McCarthy
et al. \cite{mc95}).

Hartwick (\cite{ha70}) noticed the decrease of the $B/R$ ratio in the
Milky Way as the
galactocentric radius increases.
Robertson (\cite{ro73}, \cite{ro74}), Hagen \& van den Bergh (\cite{hv74}) compared 
theoretical evolutionary tracks to observed HR diagrams (HRD) of stellar clusters 
in the Galaxy and in the Large Magellanic Cloud (LMC).
They attributed the observed 
differences between the HRD to differences in the
metal content.

\begin{table*}[htb!]
\caption{Number of supergiant stars in galactic open clusters with $\log age$ between 6.8 and 7.5. The quantity R is the galactocentric radius
and $Z$ is the
metallicity. GC means clusters toward the galactic centre, GAC toward the galactic anticentre. The identification of
supergiants is based on spectroscopic measurements. $B$ includes O, B, A supergiants and $R$ includes K, M supergiants (see text).} 
\label{tbl-1}
\begin{center}\scriptsize
\begin{tabular}{lccc|cc|ccccc}
\hline
     &	   &	       &     &	       &	 &    &    &	&    &     \\
Cluster & $\log age$ [y] & R [kpc] & $Z$ & $B$ & $R$ & $N_{O}$ & $N_{B}$ & $N_{A}$ & $N_{K}$ & $N_{M}$  \\
    &      &	       &     &	       &	 &    &    &	&    &     \\
\hline
       &     &	       &     &	       &	 &    &    &	&    &     \\
GC     &     &	       &     &         &         &    &    &	&    &	   \\     
NGC 6611	&6.88	&6.85	&0.027	&2	&0	&0	&2	&0	&0	&0\\
NGC 6604	&6.81	&6.91	&0.027	&2	&0	&1	&1	&0	&0	&0\\
Pismis 20	&6.86	&7.06	&0.026	&4	&0	&0	&4	&0	&0	&0\\
NGC 6530	&6.87	&7.18	&0.025	&1	&0	&0	&1	&0	&0	&0\\
NGC 6613	&7.22	&7.25	&0.025	&3	&0	&0	&3	&0	&0	&0\\
Trumpler 27	&7.06	&7.29	&0.025	&8	&1	&1	&7	&0	&0	&1\\
NGC 6231	&6.84	&7.32	&0.025	&3	&0	&2	&1	&0	&0	&0\\
NGC 6664	&7.16	&7.45	&0.024	&0	&1	&0	&0	&0	&1	&0\\
NGC 4755	&7.22	&7.60	&0.024	&5	&1	&0	&4	&1	&0	&1\\
NGC 6514	&7.37	&7.69	&0.023	&3	&0	&0	&1	&2	&0	&0\\
NGC 6823	&6.82	&7.71	&0.023	&1	&0	&0	&1	&0	&0	&0\\
NGC 5281	&7.15	&7.85	&0.023	&1	&0	&0	&0	&1	&0	&0\\
NGC 3603	&6.84	&7.92	&0.022	&3	&0	&2	&1	&0	&0	&0\\
IC 2944		&6.82	&7.92	&0.022	&3	&1	&1	&2	&0	&0	&1\\
NGC 3766	&7.16	&7.95	&0.022	&2	&2	&0	&2	&0	&2	&0\\
NGC 3590	&7.23	&8.05	&0.022	&1	&0	&0	&1	&0	&0	&0\\
Trumpler 18	&7.19	&8.11	&0.021	&1	&0	&0	&1	&0	&0	&0\\
Collinder 228	&6.83	&8.11	&0.021	&2	&1	&1	&1	&0	&0	&1\\
Trumpler 15	&6.93	&8.14	&0.021	&1	&1	&1	&0	&0	&0	&1\\
Bochum 10	&6.86	&8.14	&0.021	&1	&0	&0	&1	&0	&0	&0\\
NGC 6871	&6.96	&8.17	&0.021	&2	&0	&0	&2	&0	&0	&0\\
NGC 3293	&7.01	&8.18	&0.021	&2	&1	&1	&1	&0	&0	&1\\
IC 2581	&7.14	&8.23	&0.021	&2	&0	&0	&1	&1	&0	&0\\
NGC 6913	&7.11	&8.32	&0.021	&3	&0	&1	&2	&0	&0	&0\\
NGC 6910	&7.13	&8.35	&0.021	&1	&0	&0	&1	&0	&0	&0\\
Berkeley 87	&7.15	&8.37	&0.020	&0	&1	&0	&0	&0	&0	&1\\
     &     &	       &     &         &         &    &    &	&    &	   \\
GAC     &     &	       &     &         &         &    &    &	&    &	   \\
Collinder 135	&7.41	&8.62	&0.020	&0	&1	&0	&0	&0	&1	&0\\
Trumpler 37	&7.05	&8.67	&0.019	&5	&2	&0	&4	&1	&0	&2\\
Collinder 121	&7.05	&8.77	&0.019	&4	&1	&0	&4	&0	&1	&0\\
NGC 1976	&7.11	&8.83	&0.019	&0	&1	&0	&0	&0	&1	&0\\
NGC 7419	&7.28	&9.05	&0.018	&0	&5	&0	&0	&0	&0	&5\\
NGC 7235	&7.07	&9.53	&0.017	&1	&1	&0	&1	&0	&1	&0\\
NGC 2244	&6.90	&9.81	&0.016	&0	&1	&0	&0	&0	&1	&0\\
NGC 2384	&6.90	&9.86	&0.016	&1	&0	&1	&0	&0	&0	&0\\
NGC 663	&7.21	&9.86	&0.016	&7	&1	&0	&7	&0	&0	&1\\
NGC 957	&7.04	&9.89	&0.015	&1	&0	&0	&1	&0	&0	&0\\
NGC 654	&7.15	&9.91	&0.015	&1	&0	&0	&0	&1	&0	&0\\
IC 1805	&6.82	&9.92	&0.015	&1	&0	&1	&0	&0	&0	&0\\
NGC 581	&7.34	&10.00	&0.015	&1	&2	&0	&1	&0	&0	&2\\
NGC 869	&7.07	&10.07	&0.015	&6	&1	&0	&6	&0	&0	&1\\
NGC 457	&7.32	&10.13	&0.015	&1	&1	&0	&0	&1	&0	&1\\
NGC 884	&7.03	&10.29	&0.014	&4	&5	&0	&2	&2	&0	&5\\
NGC 2439	&7.25	&10.64	&0.013	&1	&1	&0	&1	&0	&0	&1\\
NGC 2414	&6.98	&10.99	&0.013	&1	&0	&0	&1	&0	&0	&0\\
Ruprecht 55	&6.85	&11.12	&0.012	&1	&0	&0	&1	&0	&0	&0\\
     &	   &	       &     &	       &	 &    &    &	&    &     \\
\hline
\end{tabular}
\end{center}

\end{table*}

Humphreys (\cite{hu79}) investigated the variation of the $B/R$ ratio in the LMC. She concluded that the radial
variation of this ratio must be weak or absent, in agreement with the study of the chemical composition of
the HII regions made by Pagel et al. (\cite{pa78}). 
In a study of the supergiants in the Galaxy and in the LMC, Humphreys \& Davidson (\cite{hd79}) 
found a decrease of the $B/R$ ratio when the metallicity
decreases. Cowley et al. (1979) examined the variation of the $B/R$ ratio across the face of the LMC and found
that it increases by a factor of 1.8 when the metallicity increases by a factor of 1.2. 

The first study of the $B/R$ ratio using stellar clusters in the
Galaxy, the LMC and the SMC (Small Magellanic Cloud) was made by Meylan \& Maeder (\cite{mm82}), 
who concluded to a steep increase of $B/R$ when the metallicity
increases, with a difference of about an order of magnitude in $B/R$ between the Galaxy and the SMC. 
Humphreys \& McElroy (1984) reexamined the $B/R$ ratio in the Galaxy, 
the LMC and the SMC, and gave new values which accounted for the incompleteness of the data. They found a $B/R$ ratio about ten times higher in
the central parts of the Galaxy than in the
SMC. 

Langer \& Maeder (\cite{lm95}) summarized some results about the values of the $B/R$
ratio and gave a $B/R$ ratio 
approximately six times larger in the solar neighborhood than in the SMC. After comparison
of different stellar models with observations, they concluded that most massive star models have
problems reproducing the decrease of $B/R$ when the metallicity decreases. Deng et al. (1996)
used the supergiant stars catalogue from Blaha \& Humphreys (\cite{bh89}) 
to make new counts in the Galaxy and in the LMC. They found a $B/R$ ratio about five
times higher in the Galaxy than in the LMC. 

To summarize the $B/R$ ratio appears to be an increasing function of metallicity. This trend is observed by
studies in the Galaxy, the Magellanic Clouds as well as in the Triangulum galaxy. 
In this work we shall take the opportunity of recent improvements brought to our knowledge of Galactic and
Magellanic clusters to reexamine carefully this question.

\section{Results from new cluster studies in the Galaxy, LMC and SMC}

\subsection{Spectroscopic counts}

We select in the Webda database (Mermilliod \cite{me95}; http://obswww.unige.ch/webda) young clusters
with $\log age$ between 6.8 and 7.5, corresponding to masses at the turn off between 8 $M_{\odot}$ and 24
$M_{\odot}$ and to  initial masses of the supergiants between about 8 and 30 $M_{\odot}$ (Meynet et al. \cite{meynet93}). Let us note
that this interval of masses for the supergiants probes a region
of the HR diagram below the observed upper limit for the
luminosity of the red supergiants, {\it i.e.} below $M_{\rm bol} \sim -9.5$
(Humphreys \& Davidson \cite{hd79}, \cite{hd84}).

We only keep the clusters with
spectroscopic measurements for the brightest stars in order to distinguish supergiants
from main sequence stars. In this way, 45 clusters are selected. For each cluster, we
count the number of supergiant stars of spectral type O, B, A, K and M.
In a few cases, the spectroscopic class for the same star is different depending
on the authors. In these cases, we adopt the most recent determination. The galactocentric
distance R of each cluster is calculated from the values of its galactic coordinates and from its distance to the
sun given by the Webda database. For each galactocentric radius, a metallicity is
associated by taking a value of $Z=0.02$ at the solar position (we take R$_{\odot}=8.5$ kpc for the solar galactocentric
distance) and an averaged radial metallicity gradient of $-0.08$ dex kpc$^{-1}$. This value has been chosen from
various studies
of the galactic gradient determined by means of spectroscopic and photometric indices of stellar populations in open clusters 
(see the discussion by Alib\'es et al. \cite{al01}).

The results are given in Table~\ref{tbl-1}, where $B$ represents the number of O, B and
A supergiants, while $R$ is the number of K and M supergiants. $N_{O}$, $N_{B}$, $N_{A}$, $N_{K}$ and $N_{M}$ indicate the
number of supergiants with spectral type O, B, A, K and M respectively. Each cluster age is
given in logarithm and taken from the Webda database. The region noted GC
refers to the region inside the solar circle and includes all the clusters with a galactocentric distance shorter
than 8.5 kpc. Likewise, the GAC region corresponds to the region outside the solar circle and includes the clusters
with a galactocentric distance higher than 8.5 kpc.

\begin{figure}[htb!]
 \resizebox{\hsize}{!}{\includegraphics{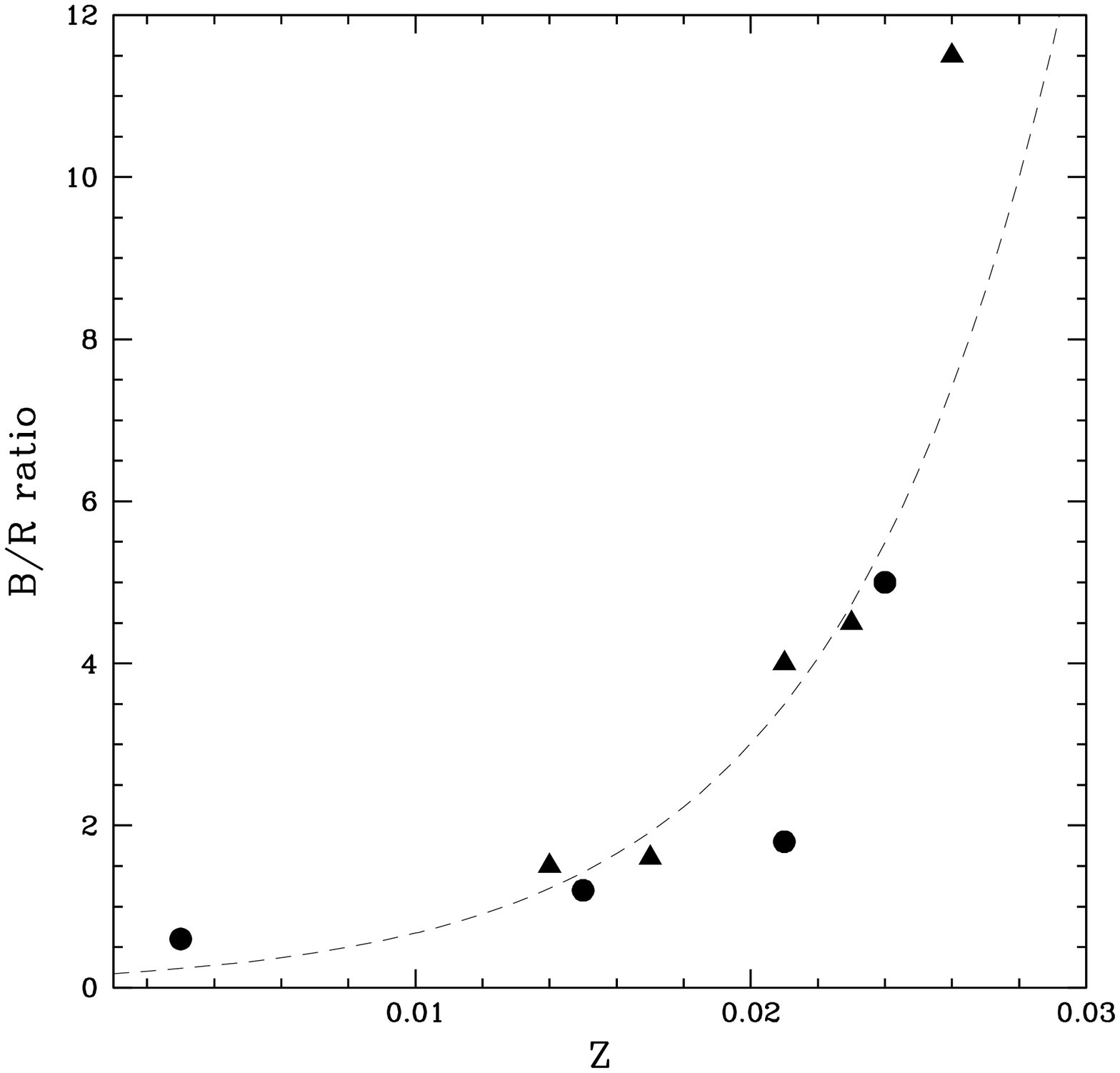}}
  \caption{$B/R$ ratio in the Galaxy and the SMC for clusters with $\log age$ between 6.8 and 7.5. The distinction between blue and red
supergiants is based on spectroscopic measurements. The triangles refer to $B$
including O, B and A supergiants and correspond to five different distance
intervals in the Galaxy (6.5--7.5, 7.5--8.0, 8.0--8.5, 8.5--10.0 and 10.0--11.5 kpc).
The dots refer to $B$ including only B supergiants; they correspond to one
value for the SMC and three
distance intervals in the Galaxy (6.5--8.0, 8.0--9.0 and 9.0--11.5 kpc). The dashed curve corresponds to the
fit for $B$ including O, B and A supergiants with $(B/R)_{\odot}=3.0$ (see text).}
  \label{fig1}
\end{figure}

\begin{figure}[htb!]
 \resizebox{\hsize}{!}{\includegraphics{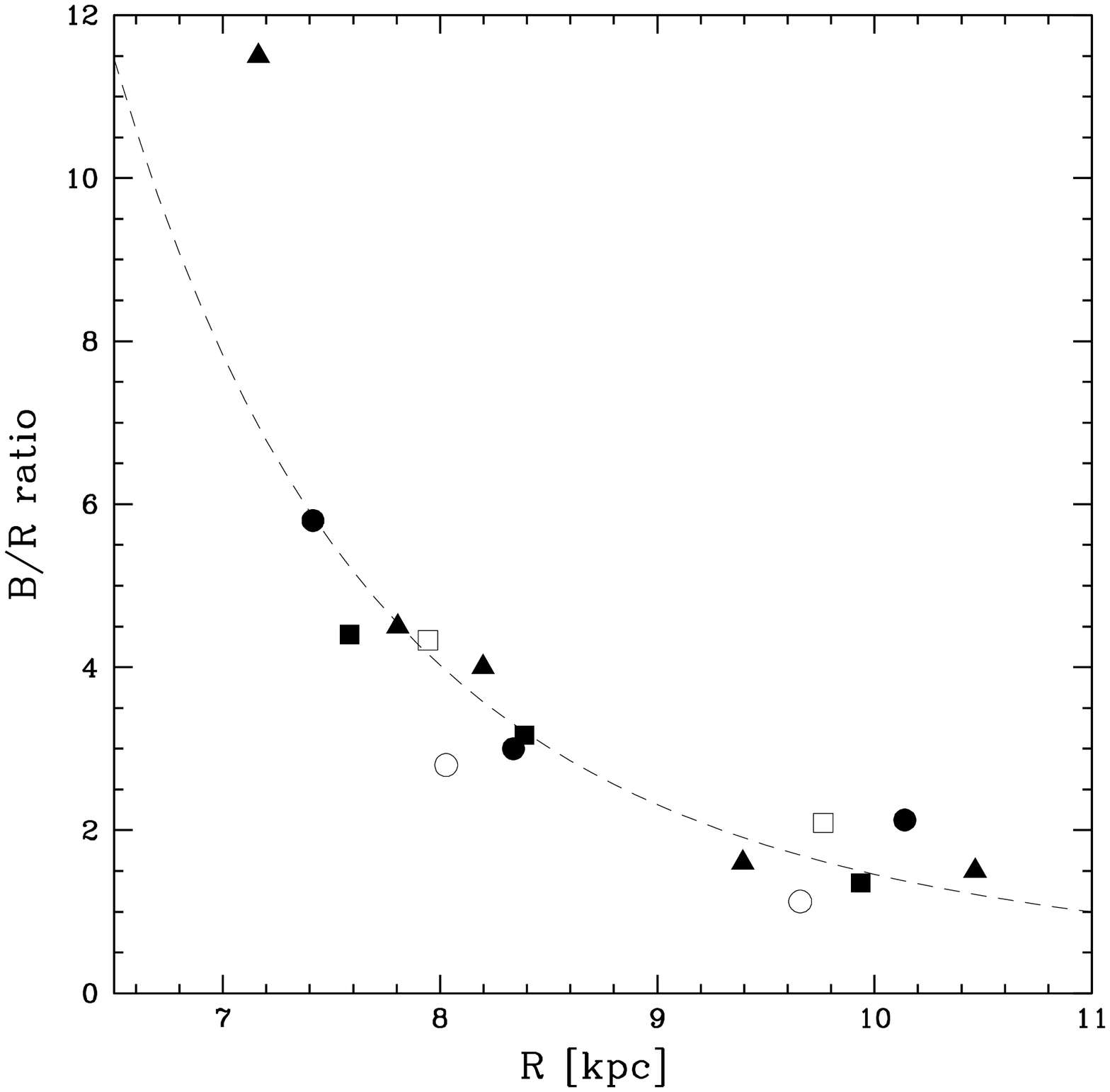}}
  \caption{$B/R$ ratio in the Galaxy for different age intervals, with distinction between blue and red
supergiants based on spectroscopic measurements.
$B$ includes O, B and A supergiants. As in Fig.~\ref{fig1}, triangles correspond to five different distance
intervals in the Galaxy (6.5--7.5, 7.5--8.0, 8.0--8.5, 8.5--10.0 and 10.0--11.5 kpc). Filled squares and circles
correspond to three distance intervals in the Galaxy (6.5--8.0, 8.0--9.0 and 9.0--11.5 kpc) and open squares and
circles correspond to the two main distance intervals GAC and GC. Triangles include clusters with
$\log age$ between 6.8 and 7.5 (i.e initial masses of the supergiants between $\sim$ 8 and 30 $M_{\odot}$). Filled circles refer to
clusters with $\log age$ between 6.8 and 7.2 (initial masses of the supergiants between 12--30 $M_{\odot}$), while filled squares include
clusters with $\log age $ between 7.0 and 7.4 (initial masses of the supergiants between 9--18 $M_{\odot}$). Open squares correspond to clusters
with $\log age$ between 6.9 and 7.1 (initial masses of the supergiants between 14--23 $M_{\odot}$) and open circles to clusters 
with $\log age$ between 7.1 and 7.3 (initial masses of the supergiants between 10--15 $M_{\odot}$). The dashed curve corresponds to the same 
fit as in Fig.~\ref{fig1}.}
  \label{fig2}
\end{figure}

On the basis of these results, we examined the variation of the $B/R$ ratio with the
galactocentric radius by grouping open clusters in different distance bins. In Fig.~\ref{fig1}, two
different binnings are shown. In the first one, clusters are grouped in three main galactocentric distance intervals 
(6.5-8.0, 8.0-9.0 and 9.0-11.5 kpc), in order to have the same number of clusters (15) in each bin. We also separate the clusters
in five distance intervals (6.5-7.5, 7.5-8.0, 8.0-8.5, 8.5-10.0 and 10.0-11.5 kpc) to check whether the variation of the $B/R$ ratio observed when considering three
intervals is still present (due to the small number of supergiants counted for each clusters, it is not pertinent
to adopt a finer binning). 
The metallicity of each bin is obtained by averaging the metallicities of all the clusters in
this bin. We precise here that the metallicities are not obtained from stars in each clusters, but from an adopted metallicity
gradient of $-0.08$ dex kpc$^{-1}$. In this way, we assign a metallicity for each distance interval mentioned above (e.g. for the GC and GAC region 
we take respectively $Z=0.023$ and
$Z=0.016$). The increase of the $B/R$ ratio with the metallicity is shown in Fig.~\ref{fig1} and in Table~\ref{tbl-3}.        

In order to obtain values of the $B/R$ ratio for lower metallicities, we select young clusters in the Magellanic
Clouds. According to a discussion by Maeder et al. (\cite{maeder99}), the average value of the metallicity
is $Z=0.007$ for the LMC and $Z=0.002$ for the SMC. 
Without sufficient spectroscopic data for the young clusters of the LMC, it is not possible
to give a reliable value for $B/R$ in the LMC. In the SMC, the only cluster satisfying the age criteria and having spectroscopic measurements is
NGC 330 ($\log age=7.00$ according to Cassatella et al. \cite{cassa96}). This cluster has been well studied and so different values exist for the $B/R$ ratio : $B/R=7/12$ (Cayrel et al.
\cite{cay88}), $B/R=9/15$ (Carney et al. \cite{car85}; Brocato \& Castellani \cite{bc92}; Bomans \& Grebel \cite{bg94}),
$B/R=12/15$ (Grebel \& Richtler \cite{gr92}). In these ratios, $B$ only includes B supergiants.
Thus, the value of the $B/R$ ratio for NGC 330 lies somewhere between 0.5 and 0.8 (when only B supergiants are counted). The value of
0.6 is currently accepted and is reported in Fig.~\ref{fig1}.

\begin{table*}
\caption{Number counts down to a limit magnitude $M_{v lim}$. The
values of $(m_{v}-M_{v})_{0}$ and $E(B-V)$ mentioned for the galactic clusters are from the Webda database
(Mermilliod \cite{me95}) and
from van den Bergh (\cite{vdb98}) for the clusters in the Magellanic Clouds. $B$ includes stars with $(B-V)_{0}\leq 0$ and $R$ stars
with $(B-V)_{0}\geq 1.2$ (see text). 
} \label{tbl-2}
\begin{center}\scriptsize
\begin{tabular}{lcccc|cc|cc|cc|cc}
\hline
	&	&	&	&	&	&	&	&	&	&	&	& \\
  &  &  &  &  & \multicolumn{2}{|c|}{$M_{v lim}=-3.25$} & \multicolumn{2}{|c|}{$M_{v lim}=-3.0$} &
    \multicolumn{2}{|c|}{$M_{v lim}=-2.5$} & \multicolumn{2}{|c}{$M_{v lim}=-2.0$} \\
 Cluster & age & $E(B-V)$ & $(m_{v}-M_{v})_{0}$ & $Z$ & $B$ & $R$ & $B$ & $R$ & $B$ & $R$ & $B$ & $R$ \\
  	&	&	&	&	&	&	&	&	&	&	&	& \\
\hline
 	&	&	&	&	&	&	&	&	&	&	&	& \\
GC	&	&	&	&	&	&	&	&	&	&	&	& \\
NGC 6611	&6.88	&0.78	&11.21	&0.027	&15	&0	&16	&0	&21	&0	&35	&0 \\
NGC 6530	&6.87	&0.33	&10.62	&0.025	&6	&0	&9	&0	&13	&0	&16	&0 \\
NGC 6231	&6.84	&0.44	&10.47	&0.025	&13	&0	&14	&0	&16	&0	&28	&0 \\
NGC 4755	&7.22	&0.39	&11.48	&0.024	&13	&1	&16	&1	&28	&1	&35	&1 \\
NGC 6823	&6.82	&0.85	&11.39	&0.023	&17	&0	&18	&0	&30	&0	&45	&0 \\
NGC 3603	&6.84	&1.34	&12.80	&0.022	&37	&2	&42	&3	&63	&3	&91	&6 \\
IC 2944	&6.82	&0.32	&11.27	&0.022	&13	&1	&16	&1	&21	&1	&25	&1 \\
NGC 3766	&7.16	&0.18	&11.21	&0.022	&5	&2	&11	&2	&16	&2	&23	&2 \\
Collinder 228	&6.83	&0.34	&11.71	&0.021	&4	&0	&7	&0	&13	&0	&18	&0 \\
NGC 6871	&6.96	&0.44	&10.99	&0.021	&15	&0	&16	&0	&20	&1	&25	&1 \\
NGC 3293	&7.01	&0.26	&11.83	&0.021	&20	&1	&21	&1	&28	&1	&33	&1 \\
IC 2581	&7.14	&0.42	&11.94	&0.021	&31	&0	&35	&0	&39	&0	&42	&1 \\
	&	&	&	&	&	&	&	&	&	&	&	& \\
GAC	&	&	&	&	&	&	&	&	&	&	&	& \\
NGC 7419	&7.28	&1.83	&10.70	&0.018	&3	&5	&4	&5	&8	&7	&20	&7 \\
NGC 7235	&7.07	&0.93	&12.25	&0.017	&19	&2	&25	&2	&36	&2	&60	&2 \\
NGC 2244	&6.90	&0.46	&10.80	&0.016	&21	&1	&23	&1	&29	&2	&32	&2 \\
NGC 663	&7.21	&0.78	&11.45	&0.016	&19	&2	&26	&2	&46	&2	&65	&2 \\
NGC 957	&7.04	&0.84	&11.29	&0.015	&5	&0	&8	&0	&14	&0	&19	&0 \\
NGC 654	&7.15	&0.87	&11.55	&0.015	&1	&0	&2	&0	&5	&0	&8	&0 \\
IC 1805	&6.82	&0.82	&11.38	&0.015	&41	&0	&46	&0	&70	&1	&96	&3 \\
NGC 869	&7.07	&0.58	&11.59	&0.015	&22	&0	&29	&0	&47	&0	&66	&0 \\
NGC 457	&7.32	&0.47	&11.93	&0.015	&8	&1	&11	&1	&14	&1	&21 	&1 \\
NGC 884	&7.03	&0.56	&11.85	&0.014	&22	&5	&24	&5	&41	&5	&51 	&5 \\
NGC 2439	&7.25	&0.41	&12.93	&0.013	&7	&1	&8	&1	&14	&1	&22	&1 \\
	&	&	&	&	&	&	&	&	&	&	&	& \\
LMC	&	&	&	&	&	&	&	&	&	&	&	& \\
NGC 2004	&6.9	&0.13	&18.50	&0.007	&29	&7	&41	&7	&58	&7	&72	&8 \\
NGC 2100	&7.2	&0.13	&18.50	&0.007	&19	&11	&20	&12	&33	&13	&39	&13 \\
NGC 1818	&7.4	&0.13	&18.50	&0.007	&12	&6	&18	&6	&32	&7	&54	&8 \\

	&	&	&	&	&	&	&	&	&	&	&	& \\
SMC	&	&	&	&	&	&	&	&	&	&	&	& \\
NGC 330	&7.0	&0.06	&18.85	&0.002	&67	&20	&79	&21	&139	&23	&211	&40 \\
 	&	&	&	&	&	&	&	&	&	&	&	& \\
\hline
\end{tabular}
\end{center}

\end{table*}

\begin{table*}
\caption{$B/R$ ratio for the galactic centre (GC), the galactic anticentre (GAC), SMC and LMC. $B$ in the column called
`Spectroscopy' only includes B supergiants, while $R$ includes K and M supergiants. In
the other columns, $B$ includes stars with $(B-V)_{0}\leq 0$ and $R$ stars
with $(B-V)_{0}\geq 1.2$ (see text).} 
\label{tbl-3}
\begin{center}\scriptsize
\begin{tabular}{lc||ccc||ccc|ccc|ccc|ccc}
\hline
	&	&	&	&	&	&	&	&	&	&	&	&	& 	&	&	&\\
    &  & \multicolumn{3}{|c||}{Spectroscopy} &\multicolumn{3}{|c|}{$M_{v lim}=-3.25$} & \multicolumn{3}{|c|}{$M_{v lim}=-3.0$} &
    \multicolumn{3}{|c|}{$M_{v lim}=-2.5$} & \multicolumn{3}{|c}{$M_{v lim}=-2.0$} \\
 & $Z$ & $B^{\dag}$ & $R$ & $B^{\dag}/R$ & $B$ & $R$ & $B/R$ & $B$ & $R$ & $B/R$ & $B$ & $R$ & $B/R$ & $B$ & $R$ &
 $B/R$ \\
 	&	&	&	&	&	&	&	&	&	&	&	&	&	&	&	&\\
\hline
	&	&	&	&	&	&	&	&	&	&	&	&	& 	&	&	& \\
GC	&0.023	&41	&10	&4.1	&189	&7	&27.0	&221	&8	&27.6	&308	&9	&34.2	&416	&13	&32 \\
GAC	&0.016	&29	&23	&1.3	&168	&17	&9.9	&206	&17	&12.1	&324	&21	&15.4	&460	&23	&20.0 \\
LMC	&0.007	&-	&-	&-	&60	&24	&2.5	&79	&25	&3.2	&123	&27	&4.6	&165	&29	&5.7 \\
SMC	&0.002	&9	&15	&0.6	&67	&20	&3.4	&79	&21	&3.8	&139	&23	&6.0	&211	&40	&5.3 \\
 	&	&	&	&	&	&	&	&	&	&	&	&	&	&	&	&\\
\hline
\multicolumn{5}{l}{$\dag$ counting only B supergiants}
\end{tabular}
\end{center}

\end{table*}

\subsection{Photometric counts}

In order to complete counts based on spectroscopic measurements, we perform counts based
on colour indices in the same way as Meylan \& Maeder (1982). To this purpose, we
select among the 45 open clusters mentioned in Table~\ref{tbl-1}, those having photometric measurements down to
$M_{v}=-2$ and with at least 15 stars brighter than this magnitude. Hence
23 clusters are selected. The distinction between blue and red supergiants is made in the following way :
when $(B-V)_{0}\leq 0$, the star is considered to have a O, B or A spectral type and is included in $B$, 
whereas for $(B-V)_{0}\geq 1.2$, the star is considered to be a K or a M and is included in $R$. This distinction remains valid for the SMC, where late-type supergiants are of earlier type than in the Galaxy
and LMC (Humphreys \cite{hub79}; Elias et al. \cite{el85}). We then count
all the stars in these two colour intervals down to a certain limit $M_{v lim}$ of the absolute magnitude. In order to be
sure that the results do not depend on the value of this arbitrary chosen limit, the
counts are made for four different limits : $M_{v lim}=-3.25$, $M_{v lim}=-3.0$, $M_{v lim}=-2.5$
and $M_{v lim}=-2.0$. The 23 selected clusters and the results are listed in Table~\ref{tbl-2}. Because
of the small number of clusters selected in this case, it is not pertinent to subdivide the clusters in several
distance intervals or age intervals, and so we consider all the clusters 
with $\log age$ between 6.8 and 7.5 and only the two main regions GC and GAC to calculate the $B/R$ ratio (see Table~\ref{tbl-3}).
 
For the Magellanic Clouds, we use the data from Elson (\cite{el91}), except for the SMC cluster NGC 330 (Vallenari \cite{va94}). 
Cluster ages are taken from Cassatella et al. (\cite{cassa96}). The same criteria of selection as those used for galactic clusters give a total of
3 clusters selected in the LMC. In the SMC, NGC 330 is the only cluster satisfying both the
criteria for age and for the number of stars brighter than $M_{v}=-2$. 
The counts are made in exactly the same way as for the Milky Way. The selected clusters and the results of the counts are given in Table~\ref{tbl-2},
while Table~\ref{tbl-3} summarizes the values of the $B/R$ ratio found in the
four main regions (GC, GAC, LMC and SMC).
Let us note here that
our selected 
clusters have a mean age approximately the same in the GC, GAC and SMC regions
(respectively 7.0, 7.1 and 7.0) which is a nice feature for the comparisons performed here.

From the results for spectroscopic and photometric counts, we find the following average relation :
$$\frac{B/R}{(B/R)_{\odot}} \cong 0.05\cdot e^{3 \frac{Z}{Z_{\odot}}}$$
where $Z_{\odot}=0.02$ and $(B/R)_{\odot}$ is the value of $B/R$ at $Z_{\odot}$ which depends on the
definition of $B$ and $R$ and on the age interval considered. For clusters with $\log age$ between 6.8 and 7.5 (initial masses
of the supergiants between $\sim 8$ and 30 $M_{\odot}$), $(B/R)_{\odot}\cong3.0$ for spectroscopic counts when $B$ includes O, B and A supergiants, and
$(B/R)_{\odot}\cong2.5$ when only B supergiants are counted. For photometric counts, $(B/R)_{\odot}$ equals
$17.2$, $18.5$, $23.0$ and $24.0$ for $M_{v lim}=-3.25$, $M_{v lim}=-3.0$, $M_{v lim}=-2.5$
and $M_{v lim}=-2.0$, respectively.

\section{Discussion of the results}

Fig.~\ref{fig1} clearly shows that the present study, based on the most recent spectroscopic
measurements found in the literature, confirms the trend already found in previous works, namely the decrease of the $B/R$ ratio with the metallicity. 
In the Galaxy, $B/R$ is approximately 3 times higher in the GC than in the GAC (Table~\ref{tbl-3}). This variation can also be observed
by separating the Galaxy in three or five galactocentric distance
intervals. Moreover, 
the $B/R$ ratio in the SMC confirms the results obtained for the Galaxy, with a value approximately 2 times
lower than in the GAC region.
Fig.~\ref{fig1} shows also that the decrease is stronger at high metallicity than at low metallicity. 
This is in good agreement with recent results from Massey (\cite{ma02}) who found a $B/R$ ratio only slightly higher
in the LMC than in the SMC.

Concerning photometric counts, we can see in Table~\ref{tbl-3} that $B/R$ is higher in the GC region than in the GAC region. The increase
of the $B/R$ ratio when the limit magnitude increases is simply due to the fact that more main sequence stars are
included in $B$. The results are stable for the limit magnitudes $-3.25$, $-3.0$, and $-2.5$, for
which $B/R$ is approximately 2.5 times higher in the GC than in the GAC (in good agreement with Meylan \& Maeder \cite{mm82}
and with the results from spectroscopic counts).
For the limit magnitude of -2.0, the value of $B/R$ for the GC seems relatively low, which likely reflects
problems of completeness.      

The $B/R$ ratios in the LMC (Table~\ref{tbl-3}) confirm the decrease of $B/R$ when the
metallicity decreases. However, the results for the SMC seem less clear. This is probably due to the fact that ratios for the SMC are based on one single cluster (NGC 330).
This is obviously insufficient to obtain precise $B/R$ ratios and thus, in this case, results must be regarded with circumspection.

What are the effects of grouping clusters of different ages on the results ? Let us recall that 
we selected clusters with $\log age$ between 6.8 and 7.5. This means that the $B/R$ ratio we obtain are for supergiants
with initial masses between about 8 and 30 $M_\odot$. Adopting narrower age intervals would give indication
on the $B/R$ ratio in a smaller range of initial masses. One can wonder to which extent the decrease of the $B/R$ ratio
with the metallicity would remain the same, if smaller age (mass) intervals were adopted.
To check this point,
we considered different age intervals 
({\it i.e.} different intervals of initial masses for the supergiants). It is obvious that a finer age interval
means less clusters selected and so a less reliable statistic. Therefore, we choose larger distance intervals when finer age
intervals are considered. First, we calculate the $B/R$ ratio in the same three distance intervals considered above (6.5-8.0, 8.0-9.0
and 9.0-11.5 kpc) for two age intervals : $\log age$ between 6.8 and 7.2 (initial masses of the supergiants between 12--30 $M_{\odot}$) and $\log age$ between 7.0
and 7.4 (initial masses of the supergiants between 9--18 $M_{\odot}$). We also choose two finer age intervals, $\log age$ between 6.9 and 7.1 (initial masses of the supergiants between 14--23 $M_{\odot}$) and $\log age$ between 7.1 and 7.3 (initial masses of the supergiants between 10--15 $M_{\odot}$), for which we calculate the $B/R$ ratio in the two
main distance intervals GC and GAC. The results of these counts are shown in Fig.~\ref{fig2}. We precise that the value of the
galactocentric radius R of each distance interval is obtained by averaging the
galacocentric radii of the clusters found in the bin.

We observe that for different age intervals, 
the $B/R$ ratio always remains higher for larger metallicities. Fig.~\ref{fig2} shows clearly that the variation of the $B/R$ ratio
with the galactocentric radius remains more or less the same whatever the interval of ages (masses) considered.
Moreover, we can see that considering finer age
intervals changes only slightly the values of the $B/R$ ratio. In that respect, the age interval between 6.8 and 7.5
appears to be a good compromise between the necessity to select a sufficiently high numbers of clusters to
have a reliable statistics, and the necessity to somewhat restrain the domain of masses.

Finally, let us stress that when comparisons with stellar evolution models are made,
the same definitions of the blue and the red supergiants 
used for obtaining the observed $B/R$ ratios has to be used to determine the theoretical
$B/R$ ratios. 
In that respect, it is worthwhile to recall here that 
even if a star is classified as a blue supergiant, it may belong
to the upper end of the Main Sequence.

\section{Conclusions}

The decrease of the $B/R$ ratio with the metallicity does not appear to be
an artifact due for instance to incompleteness but instead a robust feature
that stellar evolution models should be able to reproduce. The fact that the present day
grids of stellar models are still unable to account for this feature demands some
caution when evolutionary population synthesis models are used to interpret the integrated luminosity of
high redshift galaxies.

Recent studies by Maeder \& Meynet (\cite{mm01}) show that the inclusion in the stellar
models of the effects of rotation
changes 
the $B/R$ ratios. 
Physically, this results from the mild mixing, which leads to more helium in the region of the
H--shell burning. The opacity is lower and the intermediate convective zone less important or absent.
As convection implies a polytropic index $n=1.5$, which means a relative compactness of the internal
convective regions, the absence of an intermediate convective zone is a necessary condition for stellar expansion
to the supergiant stage.
Interestingly, rotating models 
are able to account for the numerous red supergiants seen at low metallicity, a feature that
standard models could not reproduce.

\begin{acknowledgements}
We would like to thank J.-C. Mermilliod for his help in using the Webda database.
    
\end{acknowledgements}

\end{document}